\documentclass[aps,twocolumn,superscriptaddress,nofootinbib]{revtex4-1}
\usepackage{graphicx}
\usepackage{hyperref}
\usepackage{amsmath,amssymb}

\begin{document}

\newcommand{\FIRSTAFF}{\affiliation{Department of Physics,
			University at Buffalo, SUNY
			Buffalo,
			NY 14260
			USA}}
	
\author{Wei-Chen Lin}
\email{weichenl@buffalo.edu} 
\author{William H. Kinney}
\email{whkinney@buffalo.edu} 
\FIRSTAFF
	
\date{\today}

\title{Trans-Planckian Censorship and $k$-inflation }

\begin{abstract}
We propose a more general version of the Trans-Planckian Censorship Conjecture (TCC) which can apply to  models of inflation with varying speed of sound. We find that inflation models with $c_S < 1$ are in general more strongly constrained by censorship of trans-Planckian modes than canonical inflation models, with the upper bound on the tensor/scalar ratio reduced by as much as three orders of magnitude for sound speeds consistent with bounds from data. In particular, models which satisfy the TCC, and therefore the de Sitter Swampland Conjecture, can still violate the more general condition for non-classicality of trans-Planckian modes. As a concrete example, we apply the constraint to Dirac-Born-Infeld inflation models motivated by string theory. 
\end{abstract}

%\keywords{}

\maketitle

\section{Introduction \label{sec1}}

Over the past several years, a series of so-called ``swampland'' conjectures have been proposed concerning the consistency of effective scalar field models of inflation with a UV completion in a theory of quantum gravity, including the de Sitter Swampland Conjecture and modified versions \cite{Obied:2018sgi,Garg:2018reu,Ooguri:2018wrx}. These conjectures have been widely studied, and place strong constraints on models of cosmological inflation \cite{Starobinsky:1980te,Sato:1981ds,Sato:1980yn,Kazanas:1980tx,Guth:1980zm,Linde:1981mu,Albrecht:1982wi}, in particular placing canonical single-field inflation models in tension with current observational bounds \cite{Agrawal:2018own,Kinney:2018nny}.

More recently proposed is the Trans-Planckian Censorship Conjecture (TCC), which postulates that any consistent theory of quantum gravity must forbid quantum fluctuations with wavelengths shorter than the Planck length from being redshifted by cosmological expansion to wavelengths where they become classical perturbations \cite{Bedroya:2019snp}. It is straightforward to show that if inflation continues beyond a minimal number of e-folds, modes with wavelengths on observable astrophysical scales today must have been shorter than the Planck length during inflation. This ``trans-Planckian problem'' has been well studied in the literature \cite{Martin:2000xs,Brandenberger:2000wr,Niemeyer:2000eh,Niemeyer:2001qe,Kempf:2000ac,Easther:2001fi,Hui:2001ce,Easther:2001fz}, with the generic expectation being that such modes are adiabatically rotated into a Bunch-Davies state, and have little if any effect on cosmological observables unless the scale of inflation is very high, and the scale of quantum gravity is very low \cite{Easther:2002xe}. The TCC by contrast, postulates that such modes are \textit{forbidden} in any self-consistent UV-complete theory. In this paper, we investigate the consequences of this conjecture in a general class of string-inspired inflation models. 

In canonical single-field inflation theories, quantum-to-classical ``freezeout'' of perturbations happens at approximately the Hubble length. For modes freezing out at the Hubble length, the TCC can then be formulated as 
\begin{equation}
\label{TCC_1}
\frac{a_f}{a_i}=e^{N} < \frac{H_f^{-1}}{l_P}=\frac{M_P}{H_f}, 
\end{equation}   
where the subscript $ i $ represents the onset of inflation, and the subscript $ f $ represents the moment when inflation ended in a period of reheating, initiating hot radiation-dominated expansion. The implication of the TCC on canonical-single field inflation was first studied in \cite{Bedroya:2019tba}, in which a stringent bound on the potential energy was found 
\begin{equation}
\label{TCC_Potential_1}
V^{1/4} < 6 \times 10^{8} \ \mathrm{GeV} \sim 3 \times 10^{-10} M_{P}.
\end{equation} 
This low energy scale leads to upper bounds on the first slow roll parameter, $ \epsilon <10^{-31} $, and the associated tensor/scalar ratio, $ r=16\epsilon< 10^{-30} $. Under the assumption that the TCC holds, these bounds strongly constrain the inflationary model space.\footnote{Warm Inflation \cite{Berera:2019zdd,Das:2019hto,Goswami:2019ehb} and modified initial states for perturbations \cite{Ashoorioon:2018sqb,Cai:2019hge,Brahma:2019unn} have been proposed as means to avoid these constraints.}

\section{Generalized Trans-Planckian Censorship Conjecture \label{sec2}}

The bounds (\ref{TCC_1}) and (\ref{TCC_Potential_1}) apply for the case of inflation generated by a canonical scalar field. However, in the case of a more general scalar Lagrangian with time-varying equation of state or speed of sound, the Hubble length is \textit{not} the length at which quantum fluctuations freeze out and become classical \cite{Khoury:2008wj,Geshnizjani:2011dk}. In the more general case, a general quadratic action for the curvature perturbation $\zeta$ can be written in terms of the dynamical variable $d y \equiv c_S d\tau$ as \cite{Khoury:2008wj}
\begin{equation}
S_2 = \frac{M_P^2}{2} \int{dx^3 dy q^2 \left[\left(\frac{d \zeta}{d y}\right)^2 - \left(\nabla \zeta\right)^2\right]},
\end{equation}
where $\tau$ is the conformal time, $ds^2 = a^2\left(\tau\right) \left[d \tau^2 - d {\bf x}^2\right]$, and $q$ is a function of the speed of sound $c_S$ and the slow roll parameter $\epsilon \equiv - \dot H / H^2$,
\begin{equation}
q \equiv \frac{a \sqrt{2 \epsilon}}{\sqrt{c_S}}. 
\end{equation}
Defining a canonically normalized scalar mode function $v \equiv M_{\rm P} q \zeta$, the associated mode equation becomes
\begin{equation}
v_k'' + \left(k^2 - \frac{q''}{q}\right) v_k = 0,
\end{equation}
where a prime denotes differentiation with respect to $y$. The {\it freezeout radius} $R_\zeta$ for which the quantum modes cease oscillation and become effectively classical is then
\begin{equation}
R_\zeta^{-2} = q''/q.
\end{equation}
Approximate scale invariance is obtained for $R_\zeta \simeq - y / \sqrt{2}$, which is in general dynamically completely independent of the Hubble length, and may be larger or smaller depending on the details of the dynamics. The usual limit of canonical single-field inflation can be obtained by taking $c_S = 1$ and $\epsilon \simeq {\rm const.} \ll 1$, so that
\begin{equation}
\frac{q''}{q} \simeq \frac{a''}{a} \simeq 2 a^2 H^2,
\end{equation}
so that the freezeout horizon corresponds to the comoving Hubble length $R_\zeta \simeq \left(a H\right)^{-1}$. Mode freezing occurs as long as the freezeout horizon shrinks in comoving coordinates. (This can happen even in the case of non-inflationary expansion, for example in models with $c_S > 1$ \cite{Magueijo:2008pm,Bessada:2009ns,Geshnizjani:2011dk,Geshnizjani:2014bya}.) 

\begin{figure}[h!]
	%\hspace{30px}
	\centering
	\includegraphics[scale=0.7]{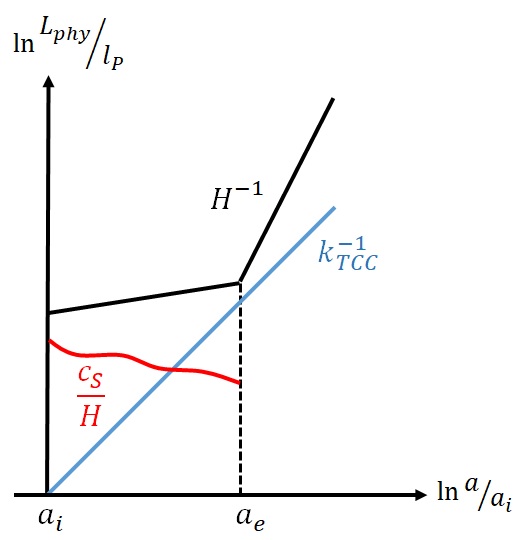}
	\caption{This diagram shows how a $k$-inflation model can satisfy Eq. (\ref{TCC_1}) but still can make trans-Planckian fluctuations classical via the crossing of acoustic horizon (red, wavy line).  In this $ \ln \frac{L_{phy}}{l_P} $ vs $\ln \frac{a}{a_i}$ diagram, the evolution of the physical length of modes is given by the family of straight lines with slope equal to 1. A mode with negative y-intercept was smaller than the Planck length at the onset of inflation. To satisfy TCC, the Hubble radius $H^{-1}$ (black) cannot intercept with the limit case $ k_{TCC}^{-1}(a_i)=l_P $. However, to satisfy GTCC, no mode originated from a length scale shorter than $ l_P $ can intercept the acoustic horizon. (Labels indicate the physical wavelength of a \textit{comoving} wavenumber $k$.)}
	\label{fig:GTCC}
\end{figure}

In this paper, we consider the generalization of the TCC to the case of non-canonical inflation models, with time-varying speed of sound $c_S$. In such models, also known generically as \textit{$k$-inflation} \cite{ArmendarizPicon:1999rj}, the freezeout length is given not by the Hubble length, but by the acoustic length,
\begin{equation}
R_\zeta \simeq \frac{c_S}{a H}.
\end{equation}
For $c_S < 1$ the acoustic horizon can be much smaller than the Hubble length, and sub-Planckian fluctuations can become classical without violating the Hubble TCC condition (\ref{TCC_1}). A schematic diagram of this possible scenario in $k$-inflation is shown in Fig. \ref{fig:GTCC}.        

In order to incorporate models with a non-canonical kinetic term, we formulate a Generalized Trans-Planckian Censorship Conjecture (GTCC) as  
\begin{equation}
\label{Most_general_GTCC}
\ln \frac{c_S(a) M_P}{H(a)} > \ln\frac{a}{a_i},
\end{equation}
where $ a_i $ is the scale factor at the beginning of the inflationary phase. For any k-inflationary model, this relation should be satisfied at any moment during the inflationary phase or sub-Planckian length scale modes would become classical via crossing of acoustic horizon. However, unlike the evolution of Hubble radius during  inflation, which can only monotonically increase within the range $ 1>d \ln H^{-1}/dN \geq0  $, the speed of sound does not have any restriction besides the range probed by Cosmic Microwave Background (CMB) non-Gaussianity measurements \cite{Akrami:2019izv}. For instance, the constraint from non-Gaussianity is $ c_S>0.079 $ ($ 95\% $, T only). However, outside of scales probed by CMB non-Gaussianity, there is considerably more freedom for variation in $c_S$, and corresponding bounds on inflation.  

A key point is that the evolution of the sound speed in non-canonical models is independent of the potential of the scalar field responsible for inflation. Therefore the dynamics of the freezout length -- here the acoustic horizon -- are likewise independent of the potential. Bedroya and Vafa \cite{Bedroya:2019snp} showed that trans-Planckian censorship in a canonical Lagrangian can be used to derive the de Sitter Swampland conjecture \cite{Garg:2018reu,Ooguri:2018wrx}. However, this linkage is not preserved for more general Lagrangians, including those typical of inflationary models in String Theory (see Sec. \ref{sec:PowerLaw}.) For $c_S < 1$, trans-Planckian censorship is always the stronger of the two constraints, in the sense that a model which satisfies the canonical version of the TCC (and therefore the de Sitter swampland conjecture), can still violate the GTCC.

\section{Simple Examples \label{sec:Three_examples}}

In this section, we consider three simple applications of the GTCC under the approximation that the Hubble length $H^{-1}$ can be taken to be approximately constant. (We discuss time-dependence of $H^{-1}$ in Sec. \ref{sec:PowerLaw}.) We further consider only the k-inflation models satisfying one of the following two conditions:\footnote{Notice that under the di Sitter limit, those two conditions on the form of the acoustic horizon reduce to conditions on the speed of sound, but generally they are not.}

\begin{itemize}
	\item The acoustic horizon decreases monotonically.
	\item  The acoustic horizon grows monotonically and at most as quickly as the wavelength of comoving wave modes, $ d \ln \left(c_S / H\right) / d N <1 $.\footnote{The reason of this condition is given in Fig. \ref{fig:excluded}.}
\end{itemize}
For models satisfying one of the above conditions, we can use the GTCC (\ref{Most_general_GTCC}) to place an upper bound on the duration of inflation by taking the end of inflation to be the earliest time at which the GTCC is saturated, so that the relation (\ref{Most_general_GTCC}) translates to an upper bound on the number of e-folds of inflation,
\begin{equation}
\label{GTCC}
N_{tot}<\ln\frac{c_S (a_e) M_P}{H_e},
\end{equation}  
where $ N_{tot}=\ln a_e/a_i $ is the total number of e-folds during inflation, while $ c_S (a_e) $ and $ H_e $ are the values of the speed of sound and Hubble parameter at the end of inflation, respectively. 

\begin{figure}[h]
	\hspace{30px}
	\centering
	\includegraphics[scale=0.45]{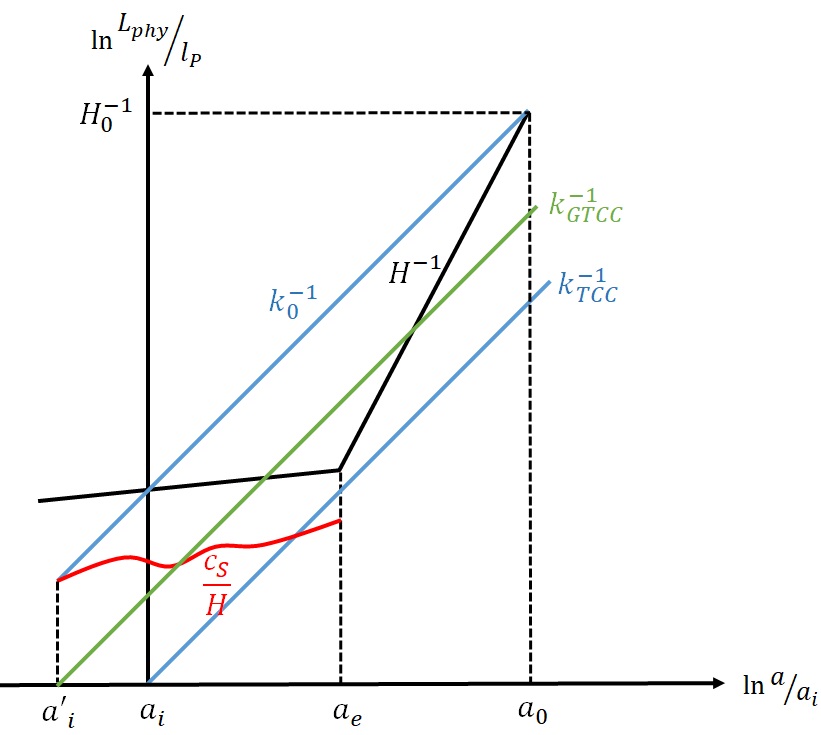}
	\caption{A schematic picture of a k-inflation model violating TCC and a canonical case with the same evolution of Hubble parameter satisfying TCC. Both of them saturate the condition of solving the horizon problem. Since the $ k_0 $ mode always exits the acoustic horizon before exits the Hubble radius, the k-inflation requires more e-folds to solve the horizon problem compared to the canonical case. As a result, the $ k_{GTCC} $ mode also exits the acoustic horizon earlier than the $k_{TCC} $ mode does. }
	\label{fig:GTCC_horizon-problem}
\end{figure}

Another issue we have to consider for k-inflation is that the condition of solving the horizon problem in k-inflation models is different from that in canonical models. This is because mode freezing happens when modes cross outside of the acoustic horizon instead of the Hubble radius, which in general happens \textit{earlier} in k-inflation than in the equivalent canonical case (Fig. \ref{fig:GTCC_horizon-problem}). If we consider a comoving wave mode with wavelength of order the horizon size today, $k_0 = (a_0 H_0)$, the condition for horizon exit of the same mode during inflation at scale factor $a_i$ is
\begin{equation}
k_0 = \frac{a_i H_i}{c_S\left(a_i\right)}.
\end{equation}
Therefore, the condition for solution of the horizon problem in a general $k$-inflation model is
\begin{equation}
\label{general_horizon_problem}
\frac{c_{S}(a_i)}{a_i H_i} \geq (a_0 H_0)^{-1}. 
\end{equation}
This reduces to the usual condition in the canonical case for $c_S = 1$.

We first consider a simple model with power-law behavior of $ c_S $:
\begin{equation}
\label{Power_law_c_S_N}
c_S(N)=c_S(a_{i})e^{s N}, 
\end{equation}
where $ s<1 $, as a pedagogical example to show how the maximum energy bound changes due to the non-trivial speed of sound. After this example, we derive the bound on the inflationary energy and the tensor/scalar ratio for models satisfying the constraint that $ c_S $ is monotonically decreasing, or monotonically increasing with $ d \ln \left(c_S / H\right) / d N <1 $. We first assume instantaneous reheating after inflation and a radiation-dominated universe throughout the rest of thermal history after reheating in Secs. \ref{subsec2.a} and \ref{subsec2.b}, which is similar to the analysis of Ref. \cite{Mizuno:2019bxy}. In Sec. \ref{subsec2.c}, we use conservation of entropy after reheating to calculate the bound without assuming a radiation-dominated post-inflation expansion, and we show that the shift in bound due to a non-thermal expansion history is small. We also generalize to the case of non-standard reheating,\footnote{The case of non-thermal post-inflationary history in the canonical case is discussed in Ref. \cite{Dhuria:2019oyf}.}  which adds logarithmic corrections. All of the results in this section are based on the same assumption made in the original paper \cite{Bedroya:2019tba}, \textit{i.e.} $ H \approx H_{inf}=const $. In Sec. \ref{sec:PowerLaw}, we then consider the more complicated case of power-law $k$-inflation with varying Hubble parameter $H\left(t\right)$, motivated by known implementations in string theory, for which the dynamics is generated by a brane evolving in a Klebanov-Strassler throat. Lastly, since we only consider the above mentioned special type of k-inflation models in the rest of the paper, we will refer Eq. (\ref{GTCC}) as the GTCC unless otherwise stated.

\subsection{ A power-law behavior of speed of sound  \label{subsec2.a}}
\begin{figure}[h]
	\hspace{30px}
	\centering
	\includegraphics[scale=0.45]{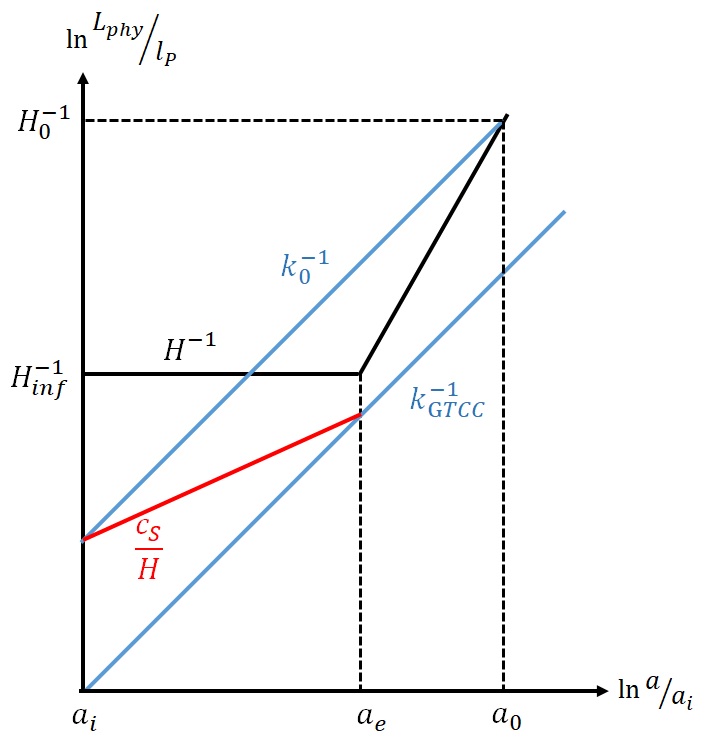}
	\caption{A schematic diagram of a $k$-inflation model with power-law increasing speed of sound: $c_S(N)=c_S(a_{i})e^{s N}$, while $ s $ is some number within the range $ 0<s<1 $. To find the upper bound on $ H_{inf} $, we saturate the bound from GTCC (\ref{GTCC}) at $ a_e $ and the bound from solving horizon problem (\ref{general_horizon_problem}) at $ a_i $.}
	\label{fig:GTCC_Power_law}
\end{figure}
In this simple example,  we consider a varying speed of sound as
\begin{equation}
\label{Hinf_power_law_cS}
c_S(N)=c_S(a_{i})e^{s N},
\end{equation} 
where $ s=const. $ is the first flow parameter of speed of sound \cite{Peiris:2007gz,Kinney:2007ag} and we only consider $ s<1 $. This includes any power-law decreasing speed of sound model ($s < 0$) and power-law increasing speed of sound model with $ d \ln c_S/dN < 1  $.\footnote{Note that we adopt the opposite sign convention for $N$ from the convention used in Refs. \cite{Peiris:2007gz,Kinney:2007ag}.} Notice that the horizon problem cannot be solved by this type of model when $ s>1 $, since the sound horizon grows more quickly than the physical wavelength of modes, which therefore never freeze out. The upper bound on the inflationary energy $ H_{inf} $ is derived by saturating the condition of solving the horizon problem (\ref{general_horizon_problem}) and GTCC (\ref{GTCC}) as shown in Fig. \ref{fig:GTCC_Power_law}.

By substituting Eq. (\ref{Hinf_power_law_cS}) into the GTCC (\ref{GTCC}), we have
\begin{equation}
\label{GTCC_power_law}
N_{tot}<\ln\frac{c_S (a_e) M_P}{H_e}=\ln\frac{c_S(a_i)  M_P}{H_{inf}} +sN_{tot},
\end{equation} 
where $ H \approx H_{inf}=const $ is used. Meanwhile, the condition of solving the horizon problem (\ref{general_horizon_problem}) under the simplification of a radiation-dominated universe (assuming instantaneous reheating) reduces to the form 
\begin{equation}
\label{LimitHorizonProblem}
\ln\frac{H_{0}^{-1}}{H_{inf}^{-1}}-2\ln c_S(a_i) \leq 2N_{tot}.
\end{equation} 
From Eqs. (\ref{GTCC_power_law}) and (\ref{LimitHorizonProblem}), we can obtain the upper bound on $ H_{inf} $ as a function of the model parameter $ s $:
\begin{equation}
\label{Limit_H_inf_power_law*}
\frac{H_{inf}}{M_P} \lesssim (10^{-20})^{\frac{3-3s}{3-s}} c_{S}(a_i)^{\frac{4-2s}{3-s}}. 
\end{equation}
where $  H_{0}=h \times 2.13 \times 10^{-42} \ \mathrm{GeV} $ with $ h=0.7 $ and $ M_P=2.435 \times 10^{18} \ \mathrm{GeV} $ are used. Notice that since we have saturated the bound from solving the horizon problem, $ c_S(a_i) $ is roughly bound by the CMB non-Gaussianity measurement and cannot be arbitrary small.    
With $ s=0 $, Eq. (\ref{Limit_H_inf_power_law*}) gives the bound in a constant speed of sound scenario as
\begin{equation}
\label{Limit_H_inf_constant}
\frac{H_{inf}}{M_P} \lesssim 10^{-20}c_{S}^{4/3}, 
\end{equation}
which reduces back to the canonical result if we further set $ c_S=1 $.

\subsection{ Bounds as functions of $ c_S(a_i) $ and  $ c_S(a_e) $ \label{subsec2.b}}

\begin{figure}[h]
	\hspace{30px}
	\centering
	\includegraphics[scale=0.45]{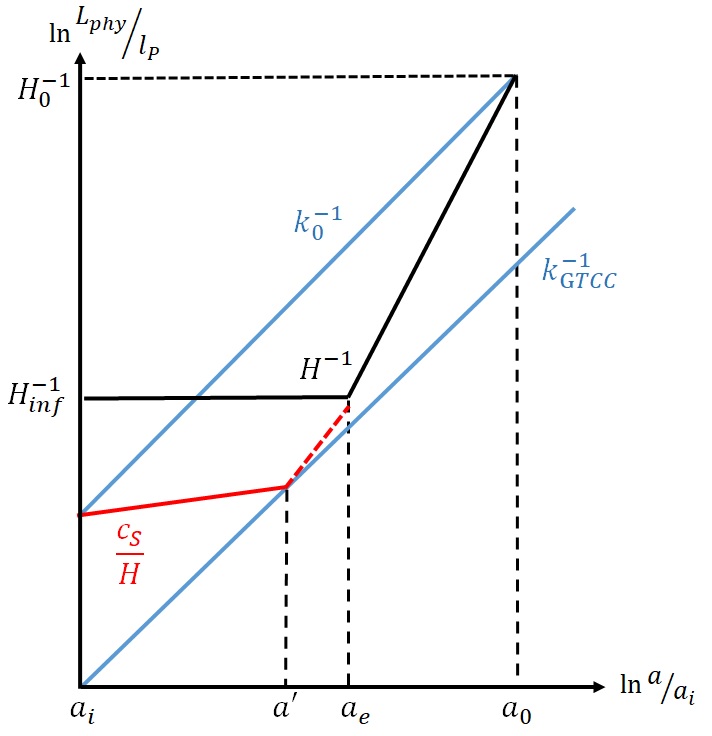}
	\caption{The GTCC (\ref{Most_general_GTCC}) cannot reduce to the condition (\ref{GTCC}) in the scenario shown here. The point where the GTCC is saturated, $ a' $, cannot be considered as the end of inflation due to a following period of $ d \ln \left(c_S / H\right) / d N <1 $. Generally speaking, the criterion is whether or not the average value $ \left[d \ln \left(c_S / H\right) / d N \right]_{\mathrm{av}} <1 $ from $ a' $ to $ a_e $. To discuss the GTCC (\ref{Most_general_GTCC}) constraint on the duration of this type of models, we need to introduce an extra model-dependent parameter, $ \ln a_e/a' $, and use $ c_S(a') $ instead of $ c_S(a_e) $. In the remainder of this paper, we restrict our discussion on models with monotonic decreasing $ c_S $ or monotonic increasing $ c_S $ with $ d \ln \left(c_S / H\right) / d N <1 $. }
	\label{fig:excluded}
\end{figure}

From the previous example it is not hard to see that the specific time dependence of $ c_S $ does not matter, as long as inflation ends whenever the first trans-Planckian mode crosses the acoustic horizon. For this type of model, the GTCC (\ref{Most_general_GTCC}) reduces to the form (\ref{GTCC}), we then can derive the upper bound on $ H_{inf} $ as a function of $ c_S(a_i) $ and  $ c_S(a_e) $. While this bound applies to a broad class of models, one can engineer exceptions by a suitable choice of time dependence for $c_S$. For example, Fig. \ref{fig:excluded} shows a counterexample that the GTCC (\ref{Most_general_GTCC}) cannot simply reduce to the condition (\ref{GTCC}) since in this scenario, the point saturating the GTCC bound (\ref{Most_general_GTCC}) cannot be considered as the end of inflation due to the following period of $ d \ln \left(c_S / H\right) / d N > 1$, in which case $a_e$ reduces to its value in the corresponding canonical inflation model. 

To find the upper bound on $ H_{inf} $ in terms of $ c_S(a_i) $ and  $ c_S(a_e) $, 
we directly use the GTCC (\ref{GTCC}) and Eq. (\ref{LimitHorizonProblem}) to obtain 
\begin{equation}
\label{Limit_H_inf_general}
\frac{1}{3}\ln\frac{M_P}{H_{0}}-\frac{2}{3}\ln [c_{S}(a_e)c_{S}(a_i)] \lesssim \ln \frac{M_P}{H_{inf}},
\end{equation}
which leads to 
\begin{equation}
\label{Limit_H_inf_general_result}
\frac{H_{inf}}{M_P} \lesssim 10^{-20} (c_{S}(a_e)c_{S}(a_i))^{2/3}, 
\end{equation}
by using $  H_{0}=h \times 2.13 \times 10^{-42} \ \mathrm{GeV} $ with $ h=0.7 $ and $ M_P=2.435 \times 10^{18} \ \mathrm{GeV} $.

We can re-derive the bound on $ H_{inf} $  (\ref{Limit_H_inf_power_law*}) for the power-law speed of sound example from Eq. (\ref{Limit_H_inf_general_result}) by the following procedure.  We first substitute the GTCC (\ref{GTCC}) into Eq. (\ref{Hinf_power_law_cS}) for $ c_S(a_e) $ to have
\begin{equation}
\label{using_formula_power_law}
c_S(a_e) <  c_S(a_i)^{1/(1-s)} \left(\frac{M_{P}}{H_{inf}}\right)^{s/(1-s)}. 
\end{equation}
Substituting Eq. (\ref{using_formula_power_law}) into Eq. (\ref{Limit_H_inf_general_result}), we again have the upper bound on the energy scale during inflation as
\begin{equation}
\label{Limit_H_inf_power_law}
\frac{H_{inf}}{M_P} \lesssim (10^{-20})^{\frac{3-3s}{3-s}} c_{S}(a_i)^{\frac{4-2s}{3-s}}. 
\end{equation}

Next, we use this result, Eq. (\ref{Limit_H_inf_general_result}), to derive the upper bound of the first slow roll parameter $\epsilon$ and the tensor/scalar ratio $r$ for $k$-inflation with $H \sim H_{inf} $. 
The power spectrum of curvature perturbations for $k$-inflation is given by 
\begin{equation}
\label{Spower_spectrum_kinflation}
\mathcal{P}_{\zeta}(k)=\frac{1}{8\pi M_P^2}\frac{H^2}{c_S \epsilon}\bigg\vert_{kc_S=H},
\end{equation}
and the power spectrum of tensor modes is 
\begin{equation}
\label{Tpower_spectrum_kinflation}
\mathcal{P}_{t}(k)=\frac{2}{\pi M_P^2}H^2\bigg\vert_{k=H}.
\end{equation}
For $k$-inflation, the tensor/scalar ratio is given by \cite{Lorenz:2008et,Powell:2008bi}
\begin{equation}
\label{T_to_S_kinflation}
r \equiv \frac{\mathcal{P}_{t}(k_*)}{\mathcal{P}_{\zeta}(k_*)} = 16 \epsilon c_S^{\left(1 + \epsilon\right)/\left(1 - \epsilon\right)} \simeq 16c_S\epsilon,
\end{equation}
with $c_S$ and $ \epsilon $ evaluated at the time when the $k_*$ mode exits the sound horizon, and approximating the Hubble parameter as constant between $k_*=H$ and $k_*c_S=H$. Substituting the upper bound Eq. (\ref{Limit_H_inf_general_result}) into Eq. (\ref{Spower_spectrum_kinflation}) with $\mathcal{P}_{\zeta}(k_*) \sim 10^{-9}$, we have the upper bound for the first Hubble slow roll parameter as
\begin{equation}
\label{epsilon_bound}   
\epsilon \lesssim 10^{-32} \frac{(c_S(a_e)c_S(a_i))^{4/3}}{c_S(a_*)} ,  
\end{equation}
where $a_*$ is the time when the pivot $k_* $ mode exits the sound horizon during inflation. From this upper bound on $\epsilon$, the upper bound for the tensor/scalar ratio is given as
\begin{equation}
\label{r_bound}   
r \lesssim 10^{-31}(c_S(a_e)c_S(a_i))^{4/3}.  
\end{equation}
Eq. (\ref{r_bound}) shows that for k-inflation, the condition that no trans-Planckian mode becomes classical by the mechanism of inflation sets a more stringent upper bound on the tensor/scalar ratio than in the equivalent canonical case.       
For example, in the simplest case that the speed of sound is also a constant during inflation, Eq. (\ref{r_bound}) reduces to
\begin{equation}
\label{r_bound_constant}   
r \lesssim 10^{-31}c_S^{8/3},  
\end{equation}
where $ c_S=1 $ reduces to the canonical result. We can give a rough estimate by using the current bound on the speed of sound from CMB non-Gaussianity, $ c_S(a_*)>0.079 $ ($ 95 \% $ from temperature data)\cite{Akrami:2019izv}, and assuming that both $ c_S(a_e), c_S(a_i) \sim 0.08 $, which lowers the upper bound on $ r $ by a factor of $  (0.08)^{8/3} \sim 10^{-3} $. Thus, the bound on the tensor/scalar ratio can be tighter by several orders of magnitude when we consider a general $k$-inflation model with $ c_S<1 $.  

For the power-law speed of sound model we considered in Sec. \ref{subsec2.a}, we can obtain the tensor/scalar ratio as a function of model parameter $ s $ by
substituting the upper bound of Eqs. (\ref{using_formula_power_law}) and (\ref{Limit_H_inf_power_law}) into
Eq. (\ref{r_bound}), which leads to
\begin{equation}
\label{r_bound_power_law}   
r \lesssim 10^{-31+\frac{80s}{3-s}}c_S(a_i)^{\frac{4(2-s)}{3-s}}.  
\end{equation}

\subsection{A more accurate estimation and including an extended reheating process \label{subsec2.c}}

To obtain a more accurate account of the evolution of universe, we can use the conservation of entropy after reheating to estimate of number of e-folds required to solve the horizon problem (see, e.g.  \cite{Bedroya:2019tba,Tenkanen:2019wsd}). However, as we will see later that the results differ at most only by one magnitude from the simplified setting we've used so far; moreover, the effect of $ c_S(a_i) $ and $ c_S(a_e) $ remains exactly the same.    

Without referring to the simplified geometrical picture as we did previously, solving the horizon problem in any k-inflation model requires that the comoving Hubble radius today fits in the comoving acoustic horizon at the beginning of inflation
\begin{equation}
\label{general_horizon_problem*}
\frac{c_{S}(a_i)}{a_i H_i} \geq (a_0 H_0)^{-1}, 
\end{equation}
which can be written as 
\begin{equation}
\label{general_horizon_problem_common}
c_{S}(a_i)e^{N_{tot}}\frac{a_{reh}}{a_e}\frac{a_0}{a_{reh}} \geq \frac{H_i}{H_0},  
\end{equation}
where $ a_{reh} $ is the scale factor at the end of reheating and the factor $ a_0/a_{reh} $ can be evaluated by the conservation of entropy after reheating 
\begin{equation}
\label{Conservation_of_entropy}
\frac{a_0}{a_{reh}}=\left(\frac{g_{*}(T_{reh})}{g_{*}(T_{0})}\right)^{1/3}\frac{T_{reh}}{T_{0}},  
\end{equation}
where $ g_{*} $ is the effective number of relativistic degrees of freedom, with $ g_{*}(T_{reh}>100\ \mathrm{GeV}) \approx 106.75 $ and $ g_{*}(T_0) \approx 3.38 $ by considering only Standard Model particle content. By assuming an instantaneous reheating immediately after inflation, \textit{i.e.} $ a_{reh}/a_e=1 $ and substituting Eq. (\ref{Conservation_of_entropy}) into Eq. (\ref{general_horizon_problem_common}), we have 
\begin{equation}
\label{horizon_S_conserved_cal_1}
\ln c_S(a_i)+N_{tot} \gtrsim 66.08 +\ln \frac{H_i}{T_{reh}},
\end{equation}
where $ \ln(T_0/H_0) \approx 67.23 $ is used. Since we assume an instantaneous reheating, we also have the relation
\begin{equation}
\label{energy_temperature_relation_Hubble}
\rho_e = 3M_P^2 H_e^2 =\frac{\pi^{2}}{30} g_{*}(T_{reh})T_{reh}^4. 
\end{equation}
Then by substituting the GTCC (\ref{GTCC}) and Eq. (\ref{energy_temperature_relation_Hubble}) into Eq. (\ref{horizon_S_conserved_cal_1}) with $ H_i \sim H_e \sim H_{inf}=const. $, we obtain a more accurate result
\begin{equation}
\label{Limit_H_inf_entropy_result}
\frac{H_{inf}}{M_P} \lesssim 10^{-19.3} (c_{S}(a_e)c_{S}(a_i))^{2/3},
\end{equation}
which only differs from the result (\ref{Limit_H_inf_general_result}) by a factor of $ 10^{-0.7} \approx 0.2 $, and the effect of speed of sound is unchanged. 

We can also consider an extended reheating process with a constant value of $ \epsilon$, which leads to the relation 
\begin{equation}
\label{reheating_relation}
\ln \frac{a_{reh}}{a_{e}}=\frac{1}{\epsilon} \ln \frac{H_e}{H_{reh}}. 
\end{equation}
To maximize the effect of this correction, we consider the limit that reheating happened just before the Big Bang nucleosynthesis, where $ T_{reh} \sim T_{BBN} \approx 4 MeV $. Then the Hubble parameter at the end of reheating is given by 
\begin{equation}
\label{BBN_temperature_relation_Hubble}
3M_P^2 H_{reh}^2 =\frac{\pi^{2}}{30} g_{*}(T_{reh} \approx 4 MeV )T_{reh}^4,  
\end{equation}
with $ g_{*}(T_{reh} \approx 4 MeV) \approx 10.75 $.
By substituting Eqs. (\ref{Conservation_of_entropy}), (\ref{reheating_relation}) and (\ref{BBN_temperature_relation_Hubble}) into Eq. (\ref{general_horizon_problem_common}) , we obtain the condition of solving the horizon problem as
\begin{equation}
\label{horizon_reheating_cal}
\ln c_S(a_i)+N_{tot} \gtrsim 114.7-\frac{95.64}{\epsilon} +\frac{\epsilon-1}{\epsilon} \ln \frac{H_{inf}}{M_P}, 
\end{equation}
where $ H_i \sim H_e \sim H_{inf}$ is used. Then by substituting GTCC (\ref{GTCC}) into Eq. (\ref{horizon_reheating_cal}), we have 
\begin{equation}
\label{Limit_H_inf_reheating_result_Log}
\ln \left[c_{S}(a_e)c_{S}(a_i)\right]-\left(114.7-\frac{95.64}{\epsilon}\right) \gtrsim \frac{2\epsilon -1}{\epsilon} \ln\frac{H_{inf}}{M_P}, 
\end{equation}
which can be written as
\begin{equation}
\label{Limit_H_inf_reheating_result}
\frac{H_{inf}}{M_P} \lesssim 10^{\frac{-49.8\epsilon +41.54}{2\epsilon -1}} (c_{S}(a_e)c_{S}(a_i))^{\frac{\epsilon}{2\epsilon -1}}. 
\end{equation}
With $ \epsilon=2 $, Eq. (\ref{Limit_H_inf_reheating_result}) reduces to result by assuming an instantaneous reheating 
\begin{equation}
\label{reheating_result_epsilon_2}
\frac{H_{inf}}{M_P} \lesssim 10^{-19.4} (c_{S}(a_e)c_{S}(a_i))^{2/3}.
\end{equation}
By substituting $ \epsilon=1 $, the upper bound of $ H_{inf} $ increases significantly with the effect of speed of sound being magnified slightly as
\begin{equation}
\label{reheating_result_epsilon_1}
\frac{H_{inf}}{M_P} \lesssim 10^{-8.3} c_{S}(a_e)c_{S}(a_i), 
\end{equation}
which reduces to the result found in \cite{Mizuno:2019bxy} when $ c_{S}(a_e)=c_{S}(a_i)=1 $ (see also \cite{Dhuria:2019oyf}).

To understand why the equation of state during reheating changes the exponent of sound speed Eq. (\ref{Limit_H_inf_reheating_result}), we can also assume an arbitrary energy scale of $ H_{reh} $ satisfying $ H_{e} > H_{reh} > H_{BBN} \approx 4 MeV  $, and we also assume $ g_{*}(T_{reh}) \approx 106.75 $ in order to simplify the discussion. Following the same procedure to obtain Eq. (\ref{Limit_H_inf_reheating_result_Log}), we then have
\begin{equation}
\label{general_Log_result}
\ln \left[c_{S}(a_e)c_{S}(a_i)\right] \gtrsim 66.7 +\frac{2\epsilon -1}{\epsilon} \ln\frac{H_{inf}}{M_P} +\frac{2-\epsilon}{2\epsilon}\ln\frac{H_{reh}}{M_P} , 
\end{equation} 
which can be written as
\begin{equation}
\label{general_Log_result_3/2}
\ln \left[c_{S}(a_e)c_{S}(a_i)\right] \gtrsim 66.7 +\frac{3}{2} \ln\frac{H_{inf}}{M_P} +\frac{2-\epsilon}{2\epsilon}\ln\frac{H_{reh}}{H_{inf}}. 
\end{equation}
Now we can see that the reason why the exponent factor changing from $ 2/3 $ in Eq. (\ref{Limit_H_inf_entropy_result}) to $\epsilon /(2\epsilon-1 )$  in Eq. (\ref{Limit_H_inf_reheating_result}) depending on the equation of state of the reheating process is due to the fact that the last term in 
Eq. (\ref{general_Log_result_3/2}) is also related to $ H_{inf} $ and can be absorbed into the form of (\ref{general_Log_result}). We also can see that if $ H_{ref} \sim H_{e} $, the last term in Eq. (\ref{general_Log_result_3/2}) is insignificant and the exponent dependence on the speed of sound is approximately $ 2/3 $.

\section{Non-Canonical Inflation in String Theory \label{sec:PowerLaw}}

In this section, we investigate how the GTCC (\ref{GTCC}) constrains the parameter space of the Dirac-Born-Infeld (DBI)-power-law inflation \cite{Silverstein:2003hf, Spalinski:2007qy, Chimento:2007es, Kinney:2007ag}, which is motivated by known implementations in string theory. We use the simplified geometrical picture, radiation-dominated after inflation, with instantaneous reheating since we have shown in Eqs. (\ref{Limit_H_inf_general_result}) and (\ref{Limit_H_inf_entropy_result}) that the effect of the speed of sound does not change under this simplification. 

We first relax the approximation that $ H \simeq \mathrm{const.} $ during inflation by considering a special class of models of $k$-inflation with a power-law background expansion. In power-law models, the Hubble parameter during inflation is given by   
\begin{equation}
\label{DBI_H_N}
H(N)=H_{i}e^{-\epsilon N},
\end{equation}
where $ \epsilon=\mathrm{const.}$ is the first Hubble slow roll parameter. Incorporating the change of Hubble parameter, $ H_i=H_e e^{\epsilon N_{tot}} $, the condition for sufficient inflation becomes  
\begin{equation}
\label{HorizonProblem_IR_POWER_LAW}
\ln\frac{H_{0}^{-1}}{H_{e}^{-1}}-2\ln c_S(a_i)<2(1-\epsilon)N_{tot}.
\end{equation} 
Saturating Eqs. (\ref{GTCC}) and  (\ref{HorizonProblem_IR_POWER_LAW}), we obtain 
\begin{equation}
\label{Limit_PowerLaw_general}
(3-2\epsilon)\ln \frac{M_P}{H_{e}}=\ln\frac{M_P}{H_{0}}-2(1-\epsilon)\ln [c_{S}(a_e)]-2 \ln[c_{S}(a_i)],
\end{equation}
which can be rewritten as
\begin{equation}
\label{Limit_PowerLaw_general_2}
\frac{H_e}{M_P}=\left(\frac{H_0}{M_P}\right)^{\frac{1}{3-2\epsilon}} \left[c_{S}(a_e)\right]^{\frac{2-2\epsilon}{3-2\epsilon}}\left[c_{S}(a_i)\right]^{\frac{2}{3-2\epsilon}}.
\end{equation}
Next, the speed of sound is given by the relation
\begin{equation}
\label{DBI_c_S_N}
c_S(N)=c_S(a_{i})e^{s N}, 
\end{equation}
where $ s=\mathrm{const.} $ Using Eq. (\ref{DBI_c_S_N}), we can write $ c_{S}(a_e) $ as
\begin{equation}
\label{DBI_c_S_end}
c_{S}(a_e)=c_{S}(a_i)e^{sN_{tot}}. 
\end{equation}
From Eqs. (\ref{GTCC}), (\ref{HorizonProblem_IR_POWER_LAW}) and (\ref{DBI_c_S_end}), we can derive the relation
\begin{equation}
\label{Limit_PowerLaw_DBI}
\frac{3-2\epsilon-s}{1-s}\ln \frac{M_P}{H_{e}}=\ln\frac{M_P}{H_{0}}-\frac{4-2\epsilon-2s}{1-s}\ln[c_{S}(a_i)],
\end{equation}
which can be rewritten as
\begin{equation}
\label{Limit_PowerLaw_DBI_2}
\frac{H_e}{M_P}=\left(\frac{H_0}{M_P}\right)^{\frac{1-s}{3-2\epsilon-s}} \left[c_{S}(a_i)\right]^{\frac{4-2\epsilon-2s}{3-2\epsilon-s}}.
\end{equation}
To apply the upper bound on $ H_e $, Eq. (\ref{Limit_PowerLaw_DBI_2}), to the DBI-Power-Law inflationary model, we can expand the exponential factors to the first order of $ \epsilon, s $, since the bound from the tilt of scalar power spectrum already limits the size of them to be small\cite{Stein:2016jja}. The first order approximation of Eq. (\ref{Limit_PowerLaw_DBI_2}) is given by
\begin{equation}
\label{Limit_PowerLaw_DBI_approx}
\frac{H_e}{M_P}=\left(\frac{H_0}{M_P}\right)^{\frac{1}{3}+\frac{2}{9}(\epsilon - s)} \left[c_{S}(a_i)\right]^{\frac{4}{3}+\frac{2}{9}(\epsilon-s)}.
\end{equation}  
Notice that since we have saturated the lower bound of $ N_{tot} $ from solving the horizon problem and both $ \epsilon, s \ll 1 $, we can use $ N_{tot}-N_{*} \sim 2.7  $ and $ N_{tot} \sim 46.2 $, which are calculated from the $ H=H_{inf} $ canonical case with the pivot scale $ k_{*}=0.05Mpc^{-1} $. Therefore, we have $ c_S(a_i) \sim c_S(a_*)e^{-2.7s} \sim c_S(a_*) >0.079 $. Substituting $ \frac{H_0}{M_P} \sim 10^{-60}$ and $ c_S(a_i) \sim 0.079 $, we obtain
\begin{equation}
\label{Limit_PowerLaw_DBI_approx_number}
\frac{H_e}{M_P}=10^{-20\left(1+\frac{2}{3}(\epsilon-s)\right)}(0.079)^{\frac{4}{3}+\frac{2}{9}(\epsilon-s)},
\end{equation}
which qualitatively shows that especially for the decreasing speed of sound case in the DBI-Power-Law models, \textit{i.e.} $ s<0 $, the constraint on $ H_e $ is more severe. 

To evaluate the bound on $\epsilon$ and $r$, We first substitute Eqs. (\ref{DBI_H_N}) into Eq. (\ref{Spower_spectrum_kinflation}) to obtain 
\begin{equation}
\label{Spower_spectrum_DBI_power}
\mathcal{P}_{\zeta}(k)=\frac{1}{8\pi M_P^2}\frac{H_e^2}{c_S \epsilon}e^{2 \epsilon N}\bigg\vert_{kc_S=H},
\end{equation}
and then use the approximation mentioned above: $  c_S(a_i) \sim c_S(a_*) >0.079 $ and  $ N_{*} \sim N_{tot}-2.7 \sim 45  $ to have
\begin{equation}
\label{epsilon_bound_DBI_PL}   
\epsilon < 10^{-32} (0.079)^{\frac{5}{3}}10^{\frac{40}{3}(2s+\epsilon)} \sim 10^{-32} (0.079)^{\frac{5}{3}}10^{\frac{80}{3}s}.  
\end{equation}
In \cite{Stein:2016jja}, the value of $ 2\epsilon+s $ is bound by the measurement of the tilt of scale power spectrum in this special model $ 0.045> 2\epsilon+s> 0.0263  $, which incorporated with Eq. (\ref{epsilon_bound_DBI_PL}) should become $ 0.045> s > 0.0263  $. This constraint on the parameter $ s $ rules out all the UV models in the DBI-Power-Law inflation. Substituting the maximal value of $ s $ into Eq. (\ref{epsilon_bound_DBI_PL}), the bound on $ \epsilon $ is
\begin{equation}
\label{epsilon_bound_DBI_IR}   
\epsilon <  10^{-32} (0.079)^{\frac{5}{3}}10^{1.2} \sim 10^{-33},  
\end{equation}
which gives a bound on the tensor/scalar ratio of $ r < 10^{-33} $. By substituting $ s=0.045 $ into Eq. (\ref{Limit_PowerLaw_DBI_approx_number}), we can also estimate the corresponding energy  
\begin{equation}
\label{Limit_PowerLaw_DBI_IR_H} 
\frac{H_e}{M_P} \lesssim 10^{-21},
\end{equation}
which is a free parameter in the original phenomenological model. This result shows that although the DBI-Power-Law models still has region in parameter space satisfying GTCC, the corresponding energy scale at the end of inflation is strongly tightened by Eq. (\ref{Limit_PowerLaw_DBI_IR_H}); meanwhile, the decreasing speed of sound scenario in this particular model is completely ruled out if we assume the GTCC.   

\section{Conclusions}
\label{sec:Conclusions}

In this paper we construct a generalization of the Trans-Planckian Censorship Conjecture to include models with variable speed of sound, since in such models mode freezing and the quantum-to-classical transition happen at the acoustic horizon rather than the Hubble length,
\begin{equation}
R_\zeta \simeq \frac{c_S}{a H}.
\end{equation}
The condition that trans-Planckian quantum modes never be redshifted into classical states is then
\begin{equation}
\ln \frac{c_S(a) M_P}{H(a)} > \ln\frac{a}{a_i},
\end{equation}
which cannot directly constrain the duration of k-inflation in the most general sense, \textit{i.e.}, more model-dependent parameters must be introduced. However, 
for models with monotonic decreasing $ c_S $ or monotonic increasing $ c_S $ with $ d \ln \left(c_S / H\right) / d N <1$, the GTCC (\ref{Most_general_GTCC}) directly gives a bound on the duration of inflation as
\begin{equation}
\label{eq:ub}
N_{tot}<\ln\frac{c_S (a_e) M_P}{H_e},
\end{equation}  
which reduces to the canonical Censorship Conjecture in the limit that $c_S = 1$. For $c_S < 1$, inflationary evolution is more tightly constrained than for the canonical limit, since the sound horizon can be several orders of magnitude smaller than the Hubble length. The upper bound on the Hubble parameter corresponding to the condition (\ref{eq:ub}) is
\begin{equation}
\frac{H_{inf}}{M_P} \sim 10^{-20} \left[c_{S}(a_e)c_{S}(a_i)\right]^{2/3},
\end{equation}
and the corresponding upper bound on the tensor/scalar ratio is
\begin{equation}  
r < 10^{-31}\left[c_S(a_e)c_S(a_i)\right]^{4/3},  
\end{equation}
We apply the generalized bound to power-law DBI inflation models, taking into account the time-evolution of the Hubble parameter, and derive an upper bound on $H$ at the end of inflation, 
\begin{equation}
\frac{H_e}{M_P} \sim \left(\frac{H_0}{M_P}\right)^{\frac{1-s}{3-2\epsilon-s}} \left[c_{S}(a_i)\right]^{\frac{4-2\epsilon-2s}{3-2\epsilon-s}}.
\end{equation}
where $\epsilon$ is the first slow roll parameter, and $c_S \propto e^{s N}$. In particular, we note that DBI inflation models with decreasing speed of sound are ruled out entirely by the GTCC. This is especially interesting, because these are models explicitly constructed in string theory -- presumably a self-consistent UV-complete theory -- which, if we forbid classicalization of trans-Planckian modes, nonetheless lie in the swampland. In this paper, we apply the general GTCC (\ref{Most_general_GTCC}) to the specific case of DBI inflation, but it would be interesting to investigate its consequences in other string-based models for inflation. 

Finally, we comment on the relation derived in Bedroya and Vafa \cite{Bedroya:2019snp} connecting  trans-Planckian censorship with the de Sitter Swampland conjecture \cite{Obied:2018sgi} bounding the slope of the scalar field potential. In the more general case considered here, the correspondence no longer holds, because the speed of sound has independent dynamics not determined by the potential. Moreover, the generalization of the de Sitter Swampland conjecture to the case of non-canonical Lagrangians is not obvious. See, for example, Ref. \cite{Mizuno:2019pcm} for a discussion of this issue. Trans-Planckian censorship, however, is a physical condition which is well-defined in non-canonical models, and any constraint arising in the non-canonical case with $c_S < 1$ is by construction \textit{more} restrictive than any corresponding constraint in the canonical limit, because the sound horizon is always smaller than the Hubble length.

\section*{Acknowledgments}
This work is supported by the National Science Foundation under grant NSF-PHY-1719690. We thank the referee for useful comments and suggestions.

\bibliographystyle{apsrev4-1}
\bibliography{TCC_DBI}

\end{document}